\documentclass[a4paper,11pt]{article}
\pdfoutput=1 % if your are submitting a pdflatex (i.e. if you have
\usepackage{jcappub} % for details on the use of the package, please
\usepackage[T1]{fontenc} % if needed
\usepackage{orcidlink}

\title{\boldmath \texorpdfstring{{\rmfamily\textsc{PatchNet}}}{PatchNet}: A hierarchical approach for neural field-level inference from \textit{Quijote} Simulations}

\author[1\, \orcidlink{0009-0009-3089-052X} ]{Anirban Bairagi}
\author[2,3,1,4\, \orcidlink{0000-0002-5854-8269}]{and Benjamin Wandelt}

\affiliation[1]{CNRS \& Sorbonne Universit\'{e}, Institut d’Astrophysique de Paris (IAP), UMR 7095, 98 bis bd Arago, F-75014 Paris, France;\;\;}
\affiliation[2]{Department of Physics and Astronomy, Johns Hopkins University, 3400 North Charles Street, Baltimore, MD 21218, USA}%
\affiliation[3]{Department of Applied Mathematics and Statistics, Johns Hopkins University, 3400 North Charles Street, Baltimore, MD 21218, USA}%
\affiliation[4]{Center for Computational Astrophysics, Flatiron Institute, 162 5th Avenue, New York, NY 10010, USA}

\emailAdd{bairagi@iap.fr}
\abstract{\textit{What is the cosmological information content of a cubic Gigaparsec of dark matter? } Extracting cosmological information from the non-linear matter distribution has high potential to tighten parameter constraints in the era of next-generation surveys such as Euclid, DESI, and the Vera Rubin Observatory. Traditional approaches relying on summary statistics like the power spectrum and bispectrum, though analytically tractable, fail to capture the full non-Gaussian and non-linear structure of the density field. Simulation-Based Inference (SBI) provides a powerful alternative by learning directly from forward-modeled simulations. In this work, we apply SBI to the \textit{Quijote} dark matter simulations and introduce a hierarchical method that integrates small-scale information from field sub-volumes or \textit{patches} with large-scale statistics such as power spectrum and bispectrum. This hybrid strategy is efficient both computationally and in terms of the amount of training data required. It overcomes the memory limitations associated with full-field training. We show that our approach enhances Fisher information relative to analytical summaries and matches that of a very different approach (wavelet-based statistics), providing evidence that we are estimating the full information content of the dark matter density field at the resolution of $\sim 7.8~\mathrm{Mpc}/h$.}

\begin{document}
\maketitle
\flushbottom

\section{Introduction}
\label{sec:intro}
The cosmological density field carries information about the key questions of modern cosmology: the origin of primordial fluctuations, the energy budget of different components of our universe, and the physical laws governing its evolution. Current and next-generation galaxy surveys such as DESI \citep{DESI2016}, SPHEREx \citep{Dore2014SPHEREx}, Euclid \citep{Laureijs2011Euclid}, and the Rubin Observatory \citep{RubinLSST2020} are mapping the large-scale structures over very large cosmological volumes at an unprecedented precision.  The cosmological information in these surveys increases over current available data sets on both large scales through larger survey volume and small scales through higher tracer density \citep{Weinberg2013}.

Cosmological inference has traditionally focused on analyzing 2-point correlations, typically the power spectrum $P(k)$ \citep{peeblesLSS, Peacock:1996ci, Tegmark2004, sanchez2008galaxy}, on large, weakly non-linear scales, where it is amenable to analytical modeling using perturbation theory (PT) \citep{Bernardeau2002, Gil_Mar_n_2012, Vlah_2016}.  However, by solely examining the power spectrum, these methods overlook valuable non-Gaussian information inherent in the dark matter distribution \citep{Carron2012, Neyrinck2009, Bernardeau2002}.

Recent analyses of BOSS (Baryon Oscillation Spectroscopic Survey) \citep{Dawson2013, Alam_2017} have revealed significant non-Gaussian cosmological information in galaxy clustering on non-linear scales. Employing higher-order clustering statistics (\textit{i.e.} bispectrum \citep{Scoccimarro2000, Yankelevich2019, Fry1984, Sefusatti2007}, trispectrum \citep{Verde_2001} etc) has yielded tighter constraints compared to using the power spectrum alone, but fully exploiting the non-Gaussian and non-linear features of the cosmological density field remains challenging. Limitations persist due to the inability of perturbation theory to accurately model dark matter clustering beyond quasi-linear scales, particularly for higher-order statistics. Moreover, these traditional approaches use a Gaussian likelihood assumption for these summary statistics in their Bayesian analysis \citep{Verde:2007wf, Leclercq:2014jda}, which may not hold true in general because of the non-linear evolution of the density field. 

One path that has been explored is to transform the density field by applying a non-linear point-wise function to it \citep{2013MNRAS.434.2961C} or by computing a "mark" \citep{infomark,hittingmark} before computing the power spectrum. This potentially moves some higher-order information of the original field into the 2-point correlation function \citep{Neyrinck_2011}. While these approaches have been shown to increase the recoverable information in the power spectrum, they explore special classes of transformations that are more limited than neural summaries.

Field-based approaches have recently gained prominence as alternatives to traditional summary statistics, aiming to directly extract cosmological information from the nonlinear matter field. The Bayesian Origin Reconstruction from Galaxies (BORG) framework \cite{Jasche2013, lavaux2019} uses Hamiltonian Monte Carlo \citep{neal2011mcmc} to jointly infer the initial conditions and the evolved large-scale structure of the Universe, incorporating observational systematics and redshift-space distortions \citep{Kaiser:1987qv}. More recently, LEFTfield \citep{Porto_2014, Stadler_2023, Tucci:2023bag} introduces a Lagrangian Effective Field Theory (LEFT)-based generative model to simulate the density field using physically interpretable perturbative expansions, allowing for fast and differentiable likelihood evaluations. These approaches benefit from their strong physical grounding and interpretability but are often limited in scale due to computational cost or rely on perturbative expansions that become inaccurate deep in the non-linear regime.

Methods such as SimBIG \citep{lemos2023simbig} pioneered retrieval of cosmological parameters from real survey data by leveraging \textit{Simulation-Based Inference }\citep{Papamakarios2019, Cranmer2020, Alsing2019} with normalizing flows or neural networks. They achieved improved constraints over traditional summaries by ingesting the observed density field \citep{Ravanbakhsh2017}. However, full-field methods face two key limitations: (i) they are computationally intensive and often infeasible on modern GPUs due to memory constraints, and (ii) they require a large number of high-fidelity simulations, which are expensive to generate. 

To address this issue, the hybrid SBI framework \citep{modi2023hybridsbiilearned} introduced the notion of dividing the observational volume into patches to address the problem of scaling the analysis to the size and resolution of current surveys. The authors find that power spectra estimated on non-linear patches suffer from supersample covariance, reducing the information efficiency except in the case of large patch sizes ($>500$ Mpc$/h$). 

The goal of our paper is to define a combination of summaries that circumvent these issues and aim at approaching the information content of the full field. 
To this end we introduce a hybrid field-based framework that combines analytical statistical summaries of the full survey, specifically the power spectrum and bispectrum, with local summaries extracted from sub-volumes (patches) of the dark matter field. These patches are chosen to be large enough to capture coherent large-scale modes in the perturbative regime within each region, yet small enough to be able to resolve non-linear structure without running into computational constraints. By integrating field-based patch summaries with traditional summary statistics applied to the entire data set, we aim to bridge the gap between large-scale, perturbative statistics and small-scale, non-linear information.

Our approach is conceptual in scope and uses idealized dark matter (DM) fields as input. Working with DM allows us to focus on raising the lower bound on the extractable cosmological information in a $(1~\mathrm{Gpc}/h)^3$ volume (down to a voxel scale of 7.8 Mpc/$h$) while avoiding the complex and unresolved problem of modeling tracer bias \citep{Desjacques2018, Eisenstein_1998}. 

As we demonstrate, this hybrid strategy significantly enhances Fisher information over $P(k)$ and $B(k)$ alone, and over the information in neural summaries trained on the limited available set of full-volume simulations. It performs on par with, or better than, recent alternatives such as the wavelet scattering transform, while remaining computationally efficient. We will find that, at a fixed voxel size, scalability of our approach to large data volumes is limited by the feasibility of computing $P(k)$ and $B(k)$, not by the memory limitations of the neural summary.

The remainder of this paper is organized as follows. Section \ref{QuijoteSims} describes our simulation suite. Section \ref{InfoCont} presents methods for measuring parameter information content in cosmological datasets. Section \ref{sec:AnalyticalSummaries} describes  the information content of analytical statistics. Section \ref{sec:NNSummaries} details our neural network methodology. In section \ref{hierarchicalinfo} we present our method of combining information from different scales. We discuss our findings and conclude in section \ref{sec:conclusion}.

\section{The Quijote Simulation Suite}
\label{QuijoteSims}
The \textit{Quijote Simulation Suite} \citep{Villaescusa-Navarro:2019bje} is a large collection of over 88,000 full $N$-body dark matter simulations designed to probe the large-scale structure (LSS) of the Universe with high statistical precision. These simulations were performed using the \texttt{TreePM} code \texttt{GADGET-III}, an enhanced version of the publicly available \texttt{GADGET-II} code \citep{Springel_2005}, and collectively span more than 40,000 distinct cosmological models. The parameter space explored includes variations in $\{\Omega_m, \Omega_b, h, n_s, \sigma_8, M_\nu, w\}$, allowing for a comprehensive study of cosmological dependencies in the LSS.

Each simulation models the evolution of dark matter within a cubic volume of $1\,(\mathrm{Gpc}/h)^3$, with outputs available at three spatial resolutions: $256^3$, $512^3$, and $1024^3$ grid points. This enables analyses at various levels of detail, depending on the specific requirements of the application.

The suite includes multiple targeted subsets designed for specific inference tasks. Notably, the \textit{Latin Hypercube (LH)} subset comprises 2,000 simulations that uniformly sample five cosmological parameters—$\{\Omega_m, \Omega_b, h, n_s, \sigma_8\}$—within prior ranges centered around the \textit{Planck 2018} best-fit values \citep{Planck2018}. This subset is particularly suited for machine learning–based emulation and parameter inference. In addition, the suite offers 15,000 simulations at a fixed \textit{fiducial} cosmology of $\Omega_m=0.3175,$ $\Omega_b=0.049,$ $h=0.6711,$ $n_s=0.96244$, and $\sigma_8=0.834$ for precision estimation of statistical observables, and 500 \textit{paired} simulations generated via finite differences along each parameter direction. These are essential for robust derivative estimation and \textit{Fisher matrix} analyses \citep{Alsing:2017var, Jung:2024esv}.

In this work, we utilize the \textit{real-space dark matter density field} at redshift $z = 0$, downsampled to $128^3$ grid. This provides a computationally efficient yet sufficiently detailed representation of the density field, suitable for hierarchical inference using neural networks. 

\section{Measuring Information: Fisher Information Formalism}
\label{InfoCont}
In order to extract the maximum amount of information from observational or simulated data $d(\theta)$, it is crucial to  quantify how informative the data is about the underlying cosmological parameters $\theta$. A widely used approach is the \textit{Fisher information matrix} \citep{1997ApJ...480...22T,Tegmark_1997,Alsing:2017var, Heavens2000}, which provides a lower bound on the variance of any unbiased estimator via the \textit{Cramér-Rao inequality} \citep{Rao1945, Cramer1946, Tegmark_1997, Alsing:2017var}-
\begin{equation}
    \mathrm{Var}(\theta) \geq \left(F^{-1}\right)_{\theta\theta},
    \label{infoineq}
\end{equation}
where the Fisher matrix $F$ is defined as
\begin{equation}
F = -\left\langle \nabla_\theta \nabla_\theta^T \mathcal{L} \right\rangle = \left\langle \nabla_\theta \mathcal{L} \nabla_\theta^T \mathcal{L} \right\rangle,
\end{equation}
and $\mathcal{L} = \log \mathcal{P}(d|\theta)$ is the log-likelihood of the data given the parameters. The expectation value is taken over different realizations of the data $d$ at fixed parameters $\theta$. In the information inequality Eq.~\ref{infoineq}, the Fisher matrix is evaluated at the fiducial value $\theta = \theta_\star$.

When the likelihood is approximately Gaussian, the log-likelihood takes the form
\begin{equation}
    \mathcal{L} = -\frac{1}{2}(d - \mu(\theta))^T C^{-1} (d - \mu(\theta)) - \frac{1}{2} \log |2\pi C|,
\end{equation}
where $\mu(\theta)$ is the mean prediction for the summary statistic at parameters $\theta$, and $C$ is the covariance matrix.  This leads to the simplified expression for the Fisher information matrix \citep{Alsing:2017var}:
\begin{equation}
    F = \nabla_\theta \mu^T \, C^{-1} \, \nabla_\theta \mu.
    \label{FIM}
\end{equation}

This expression shows that the information content of a statistic depends on both its sensitivity to the parameters and the covariance structure of the data. Accurate estimation of both the mean response $\nabla_\theta\mu$ and the covariance matrix $C$ in Eq.~\ref{FIM} critically requires a sufficient number of simulations \citep{Coulton-Wandelt2023}. Given the high information content of the dark matter clustering statistics employed in this work, the number of simulations available in the \textit{Quijote} suite (see Section \ref{QuijoteSims}) is sufficient to compute accurate finite-difference approximations of the parameter derivatives and covariance in Eq.~\ref{FIM}. 

Crucially, the data vector $d$ need not be limited to standard observables such as the power spectrum or bispectrum.  As long as the conditions leading to Eq.~\ref{FIM} are satisfied, any statistic can serve as the input to the Fisher analysis.  In the subsequent sections, we explore a variety of such summaries derived from the \textit{Quijote} dark matter overdensity field — ranging from traditional power spectrum measurements to neural-network-based representations and wavelet scattering coefficients to identify the most informative and optimally compressed summary for cosmological inference.

In our analysis, the Fisher information matrix of any statistic is computed using 5,000 of the suite’s fiducial simulations for the covariance matrix $C$ and 500 pairs of finite difference simulations for the derivative terms $\nabla_\theta \mu$. For simplicity, we show parameter constraints only for $\{\Omega_m,\sigma_8\}$ though our analysis has been done on all five parameters unless otherwise specified. This framework provides a benchmark against which we can evaluate the performance of neural network-based inference methods, which we discuss in the following sections.

\section{Analytical Summary Statistics and Information Content}
\label{sec:AnalyticalSummaries}

In this section we define the analytical summary statistics that will be combined or compared with the neural summaries discussed in the following section. 

\subsection{Power spectrum and Bispectrum}
The matter power spectrum $P(k)$ is one of the most fundamental statistical tools in cosmology, capturing the variance of matter density fluctuations as a function of scale. It quantifies the two-point correlation structure of the density field in Fourier space
\begin{equation}
    \langle\delta(\boldsymbol{k})\delta(\boldsymbol{k^\prime})\rangle = (2\pi)^3 \delta^3_D(\boldsymbol{k} + \boldsymbol{k^\prime}) P(k),
\end{equation}
where $\delta_D$ is the Dirac delta function. $P(k)$ provides key insights into the amplitude and scale-dependence of matter clustering and has been instrumental in constraining cosmological parameters from large-scale structure surveys~\cite{Tegmark2004, sanchez2008galaxy, Planck2018}.

However, the power spectrum captures only pairwise correlations. To access higher-order information, the bispectrum $B(k)$—the Fourier counterpart of the three-point correlation function—is employed
\begin{equation}
    \langle \delta(\boldsymbol{k}_1)\delta(\boldsymbol{k}_2)\delta(\boldsymbol{k}_3) \rangle = (2\pi)^3 \delta^3_D(\boldsymbol{k}_1 + \boldsymbol{k}_2 + \boldsymbol{k}_3) B(k_1, k_2, k_3).
\end{equation}
The dark matter bispectrum probes nonlinear gravitational evolution, offering additional sensitivity to cosmological parameters beyond $P(k)$ alone \citep{Scoccimarro2000, Sefusatti2006, Yankelevich2019, DAmico:2022osl}. 

These summaries (and their spherical analogs) have formed the basis of parameter inference in several landmark cosmological analyses, including studies of the cosmic microwave background \citep{Planck2018, SDSS-WMAP:Tegmark2004}, weak lensing \citep{Mandelbaum_2013,KiDS-1000}, galaxy redshift surveys \citep{SDSS:Tegmark2006, DES:2022, Ivanov_2023}. To benchmark the information content of these traditional summaries on dark matter simulations, we compute the Fisher information matrix (Eq.~\ref{FIM}) for the \textit{Quijote} simulation suite \cite{Villaescusa-Navarro:2019bje}. We generate $128$ linearly binned $P(k)$ for each realization using \texttt{Powerbox} and the bispectrum in the Quijote simulation suite truncated at  $k_{max}=0.4 \ h/\text{Mpc}$ used in our data vectors. The covariance matrix is estimated from $5,000$ fiducial simulations and parameter derivatives computed via 500 finite-difference pairs. 

\begin{figure}[htbp]
    \centering
    \includegraphics[width=0.5\linewidth]{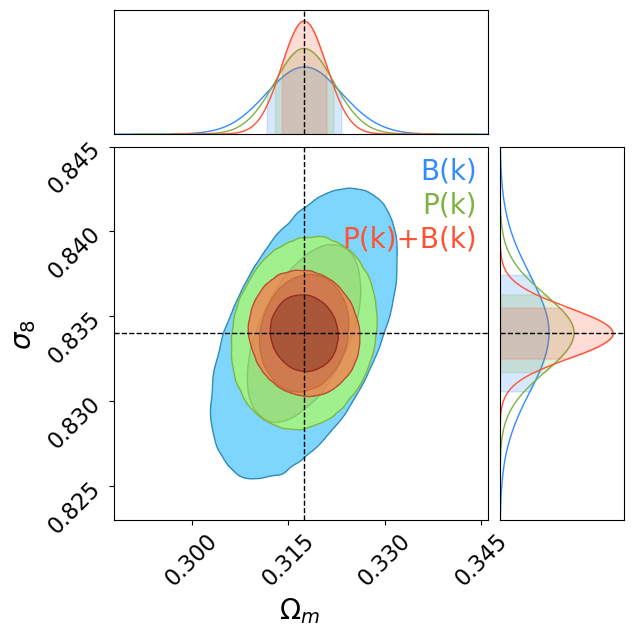}\includegraphics[width=0.5\textwidth]{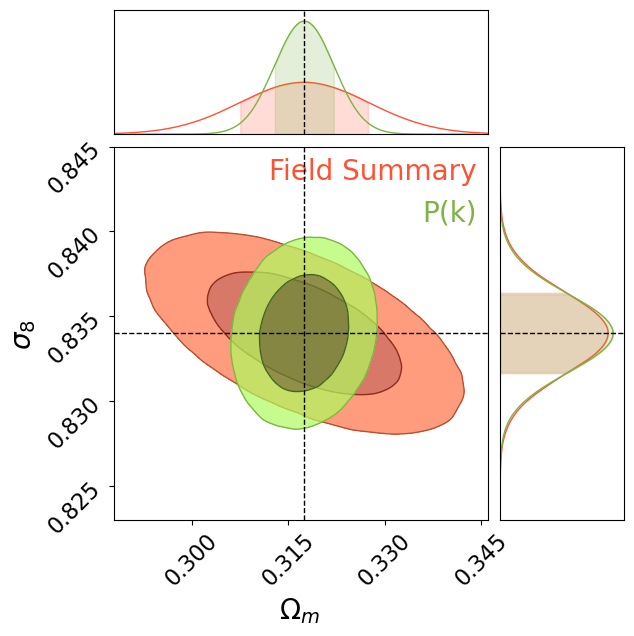}
    \caption{Left: Fisher information in traditional summary statistics. A combination of $P(k)$ and $B(k)$ contain more information than these summaries alone. Right: Information content in field level summary vs $P(k)$. A  full-field neural summary from the CNN (see section \ref{sec:fullfield}) trained on the $(1~\mathrm{Gpc}/h)^3$ dark matter density field fails; it gives worse constraints even than the power spectrum alone, particularly for  $\Omega_m$. To reach optimality would likely require both a more complex network architecture and a far larger number of training simulations. }
    \label{fig:PkBkField}
\end{figure}

Fig.~\ref{fig:PkBkField} shows the resulting marginalized uncertainties in the form of a corner plot, generated using \texttt{ChainConsumer} \cite{Hinton2016}. 
We see that the combined summary of \textit{Quijote} matter $P(k)$ and $B(k)$ increases the information content, Eq.~\ref{FIM}, over that from the power spectrum alone. These results serve as a baseline for comparison with field-based or machine-learned summaries, which aim to extract more of the information content present in the nonlinear regime \citep{Carron_2013, Hahn_2021}.
 
\subsection{Wavelet Scattering Transform}
The Wavelet Scattering Transform (WST) \citep{Mallat2012, bruna2012invariantscatteringconvolutionnetworks} has recently emerged as a powerful tool for extracting hierarchical features from large-scale structure (LSS) data while preserving statistical properties relevant for cosmological inference \citep{Cheng2020, Valogiannis_2022, eickenberg2022waveletmomentscosmologicalparameter}. Unlike traditional summary statistics such as the power spectrum or bispectrum, WST captures non-Gaussian information from the matter density field with a stable, interpretable, and computationally efficient representation.

We compute a feature vector based on the WST of the dark matter density field and measure its information content as a point of comparison to the trained neural approaches we will discuss in the following. The details of our WST implementation are discussed in Appendix \ref{Appendix:WST}.

\section{Neural estimators of the cosmological parameters and its information content}
\label{sec:NNSummaries}
In this section, we define and discuss the neural approaches to extract informative summaries from the \textit{Quijote} dark matter density field, including the implementational details to ensure reproducible research. We discuss the further details on the exploration of architectures and training procedures that led both the full field CNN and \textsc{PatchNet} in Appendix \ref{app:NN}.

\subsection{Full Field}\label{sec:fullfield}
% Any field-based analysis comes with a lot of complexity due to high high-dimensional complex structural form. 
We employ a 3D Convolutional Neural Networks (CNN) to perform a field level analysis on the Dark Matter (DM) density fields. CNNs consist of several convolutional layers of optimized filters that are convolved across the input to extract features in a hierarchical scheme. CNNs were primarily developed to perform image based tasks and later introduced in different domains because of its ability to (1) provide hierarchical representations directly from the raw data and extract increasingly complex features and semantic information by combining low-level features from previous layers hierarchically, (2) extract effective feature extraction regardless of their position in the input data due to translational invariance through convolutional operations, and (3) exploit local receptive fields. \\

To infer $\Lambda$CDM cosmological parameters $\{\Omega_m, \sigma_8\}$, we train a 3D CNN on the \textit{Quijote Latin Hypercube} that takes the DM over-density fields($\delta$) as the input and predicts the parameters $\hat{\theta}$. To improve NN training convergence, the over-density fields were normalized by the average standard deviation of all the fiducial simulations and we apply a log-transformation $\ln(2+\delta_\text{norm})$ to the field before feeding it into the network. 

Our CNN model contains three convolutional blocks consisting of a $4\times4\times4$ kernel convolution and average pooling layer followed by two fully connected layers accompanied with the nonlinear activation LeakyReLU(0.5). 

\begin{figure}[htbp]
    \centering
    \includegraphics[width=6.in]{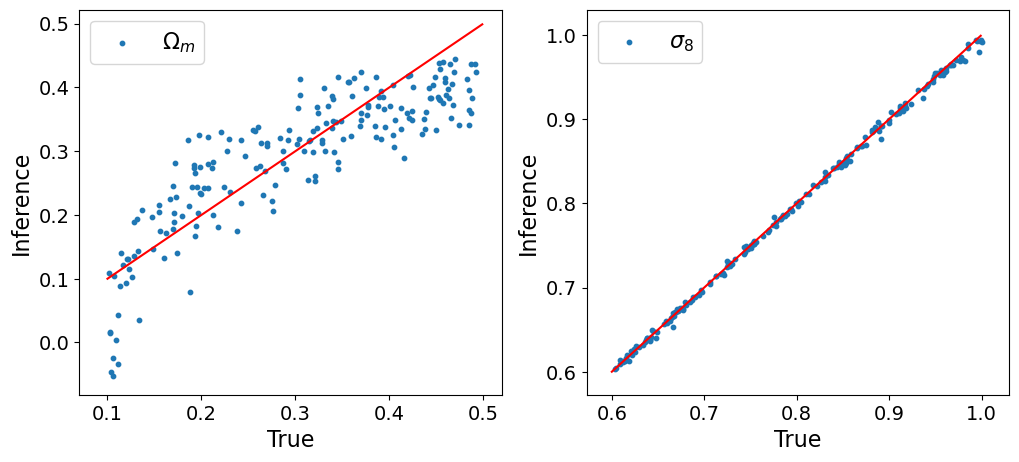}
    \caption{
   Inferred cosmology from the full-field CNN vs corresponding true cosmology computed on held out test data. The full field CNN acts on the entire $(1~\mathrm{Gpc}/h)^3$ density field sampled on a grid of $1024^3$ voxels. While the model produces accurate estimates of the parameters everywhere except near the  edge of the prior, the inference is suboptimal, as shown in Fig.~\ref{fig:PkBkField}.}
    \label{fig:field_truevsinference}
\end{figure}

For CNN training, we divide the 2,000 \textit{Latin Hypercube}  simulations into training (1500), validation (300), and test (200) sets. We use batches of 16 realizations at a time to fit it on the GPU during training. Adam optimizer \cite{kingma2017adammethodstochasticoptimization} with an initial learning rate of 0.001 and momentum($\beta$) of 0.9 has been used to minimize the log MSE loss $L$ between the true parameter $\theta$ and inferred parameter $\hat{\theta}$
\begin{equation}\label{logloss}
    L=\sum_i \ln (MSE)_i=\sum_i \ln \left(\frac{1}{N_{batch}}\sum_{n=0}^{N_{batch}} \big[\theta_i^{(n)}-\hat{\theta}_i^{(n)}\big]^2\right)
\end{equation}
where $i$ denotes different parameter classes. The learning rate is reduced by $\gamma$=0.9 after each 2 epochs using pytorch stepLR. On a single V100 GPU, NN training on 1500 Latin Hypercube DM density fields took nearly 3 hours for 200 epochs and model weights with lowest validation loss were saved for inference. We use the trained model to infer cosmology from the test data and plot the predicted cosmology from the density field for the corresponding true parameters in Fig.~\ref{fig:field_truevsinference}. 

% \begin{figure}[htbp]
%     \centering
%     \includegraphics[width=4.35in]{fig/fieldvsPk.png}
%     \caption{Fisher from field level summary \bdw{The point of this figure is to show that na\"ive training on the full field doesn't even extract as much information as the power spectrum. You had a figure like that in your slides that you can use here.  }}
%     \label{fig:timeseries_para}
% \end{figure}
In the Fig.~\ref{fig:PkBkField} the field-based summary is clearly suboptimal: its Fisher information is smaller (\textit{i.e.}, the contours are larger) than that computed from the power spectrum alone $P(k)$. 

One issue is not having sufficient training simulations required for obtaining optimal neural estimators. In \citep{Bairagi:2025sux} we showed  the importance of having a sufficient number of training simulations in order to get an optimally informative summary. 

Beyond training set size, architecture search and hyperparameter tuning for the field-level analysis are non-trivial. Cosmological fields have complex non-linear structure. The detailed structure depends on the initial conditions; the field configurations vary even if the cosmology is fixed. 

\subsection{\textsc{PatchNet}}
We will now address these issues. Rather than generating a large number of monolithic simulations and searching for optimal neural architectures, we propose a solution that exploits the cosmological scale hierarchy. 

\subsubsection{Motivation}
The isotropic and homogeneous, (nearly) Gaussian cosmological seed perturbations have small amplitude in the early universe and remain linear on large scales. The power spectrum is known to be a sufficient statistic in this regime. On intermediate and smaller scales, non-linear evolution leads to the transfer of the information to higher-order $n$-point function and the emergence of cosmic web features (nodes, filaments, sheets, voids). The information in these features is no longer fully represented in terms of simple correlation functions. 

Our key idea is to use the power spectrum and the bispectrum in the large and intermediate scale regime where they are optimal and to supplement them with neural network summaries that are trained in the small-scale regime where they are most useful.

We call the implementation of this idea  \textsc{PatchNet}. To extract the small-scale information, we split the field into smaller patches and train the neural network to extract the cosmological information from each patch. We can then aggregate the information from these patches and combine with the power spectrum.

\subsubsection{Advantages of \textsc{PatchNet}}
This approach has multiple advantages for simulation, training, and inference: 1) each simulation generates a large number of patches, increasing the effective training set size for the network; 2) the training simulations do not need to be of the same size as the full survey data set, reducing computational cost and memory use,  3) the neural network is smaller and has a smaller number of trainable parameters; 4) the size of the input to the network is much smaller, avoiding the necessity to parallelize across GPUs due to memory limitations, and 5) the network does not have to be trained to extract global Fourier-space features such as the power spectrum which are, in any case, part of the standard tool set of cosmological summary statistics.

% And for performing SBI we will be needing a training set of many such large high resolution cosmological simulations which are far too costly to generate. 
\subsubsection{Implementation}
\textsc{PatchNet} is a 3D CNN  applied to subvolumes (or patches) of the full field and infers the cosmology of the each patches based on these small scale structures. We create 512 patches of $(125\ \mathrm{Mpc}/h)^3$, sampled on a $16^3$ grid from each realization of $(1~\mathrm{Gpc}/h)^3$ \textit{Quijote} DM fields. 

The patch size of 125 Mpc$/h$, corresponding to $k_\text{patch}= 0.05 h/$Mpc , was chosen to balance computational feasibility (which favors smaller patch sizes) with the requirement to connect seamlessly to the perturbative regime where the power spectrum and bispectrum extract the full information content. 

Before training, the patches were normalized using the average standard deviation of the full fields of the fiducial simulations. This CNN consists of three convolutional blocks made up of $3\times3\times3$ convolution and average pooling followed by four fully connected layers. We have used LeakyReLU(0.5) activation after each of the full connected layers except the final layer.
\begin{figure}[htbp]
    \centering
    \includegraphics[width=\linewidth]{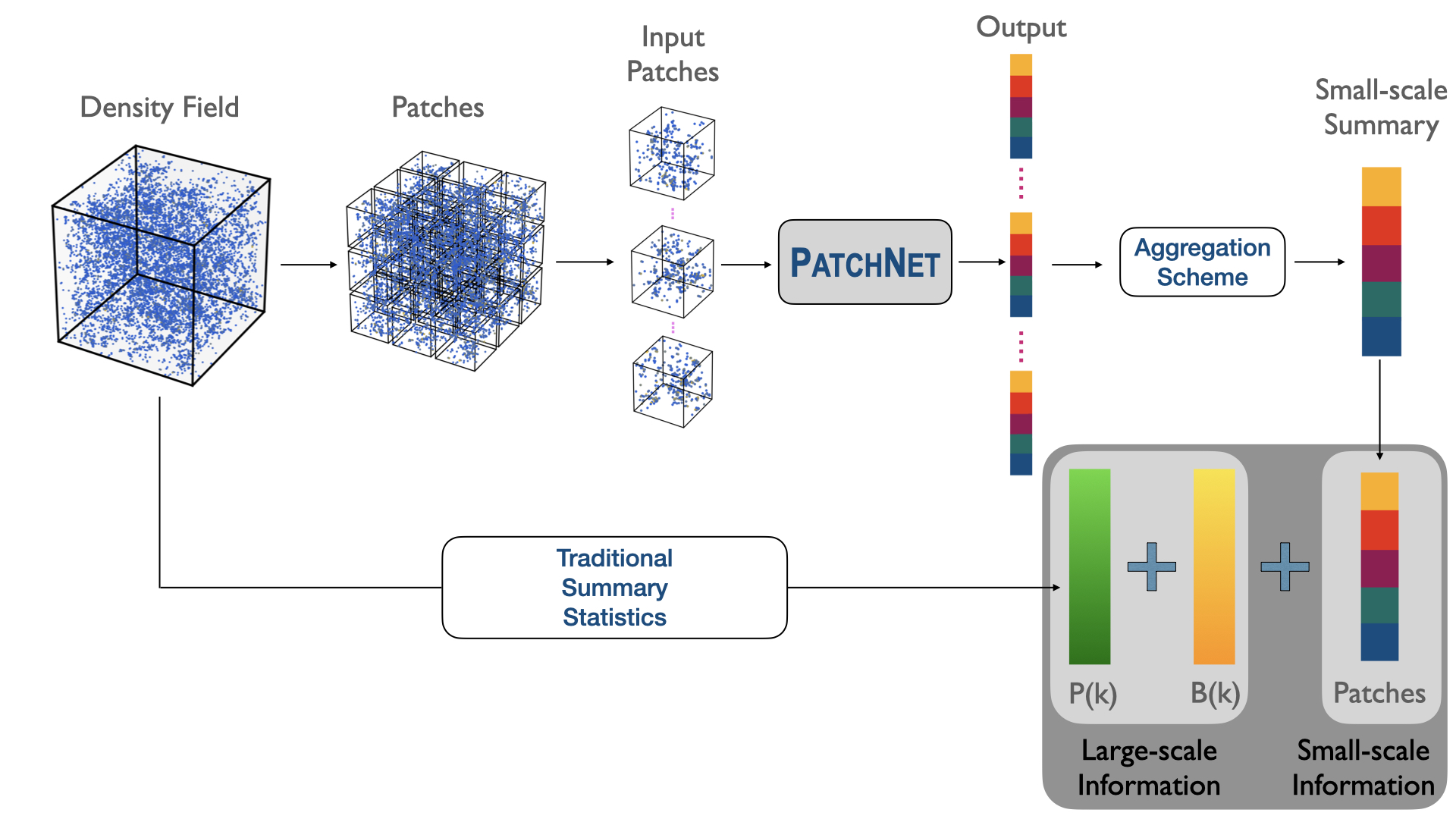}
    \caption{\textsc{PatchNet} framework: The large density field is divided into smaller patches of sufficient size to cover the perturbative scales. Then, these patches are fed into a 3D CNN architecture called \textsc{PatchNet} to constrain cosmology from the small-scale structures. We take the mean aggregate of the NN output to get the small-scale summary, which is then concatenated with large-scale statistics given by $P(k)$ and $B(k)$.}
    \label{patchnetfig}
\end{figure}
We minimize the loss function 
\begin{equation}\label{patchloss}
    L=\sum_i \ln (MSE)_i=\sum_i \ln \left(\frac{1}{N_{batch}\times N_{patch}}\sum_{n=0}^{N_{batch}} \sum_{p=0}^{N_{patch}} \big[\theta_i^{(n,p)}-\hat{\theta}_i^{(n,p)}\big]^2\right)
\end{equation}
at the patch level and optimize the model weights using Adam \citep{kingma2017adammethodstochasticoptimization} with momentum $\beta=0.9$ and learning rate starting with $0.001$ and then reduced by a factor $\gamma=0.9$ every 10 epochs.

\begin{figure}[htbp]
    \centering
    \includegraphics[width=\linewidth]{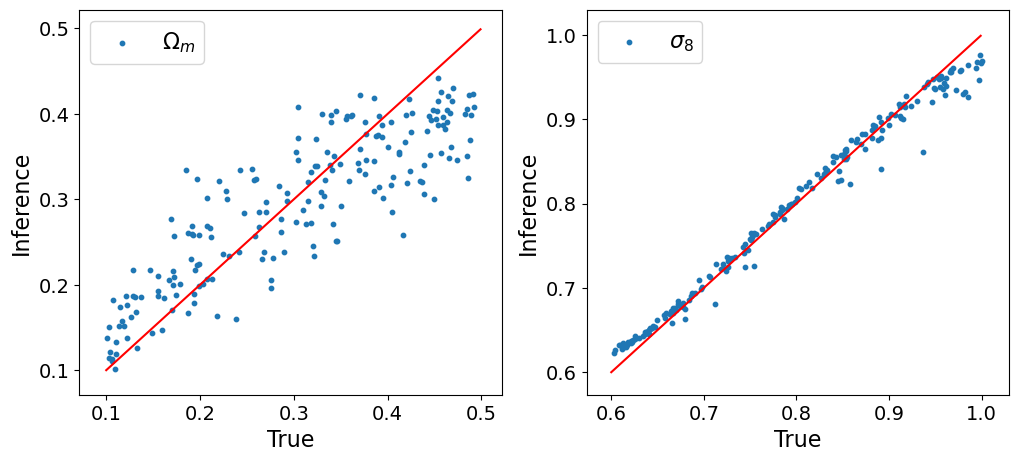}
    \caption{Inferred cosmology from aggregated \textsc{PatchNet} summary vs corresponding true cosmology computed on held out test data. The model predicts the parameters from $(125~\mathrm{Mpc}/h)^3$ dark matter patches of $16^3$ resolution. The parameter estimates of these test simulations are based on the mean value of predicted parameters across 4096 patches from each realization. It has access to small scales information only, while  the full field neural summary Fig.~\ref{fig:field_truevsinference} contains information from all scales. Patches outperform combining with the large-scale information in $P(k)$ and $B(k)$ as shown in Fig.~\ref{fig:patchvswst}. }
    \label{fig:patch_truevsinference}
\end{figure}

We train this network with a batchsize of 32 and 64 patches from each realization on a single V100 GPU for 500 epochs which required around 4.5 hours. The trained model was saved based on validation loss like earlier. Once trained for 5 parameters we use the mean value of the target parameters ($\hat{\theta}:\{\Omega_m, \sigma_8\}$) from patches as our predicted output and use these in our further analysis. We apply the trained model on the test data for parameter inference from small scales and show its accuracy with respect to the true labels in Fig.~\ref{fig:patch_truevsinference}. While patch-based summaries primarily capture information from small-scale modes, Fig.~\ref{fig:field_truevsinference} includes contributions from all scales. The patch-based summaries yield significantly tighter overall constraints when complemented by large-scale statistics such as the power spectrum and bispectrum.

\subsubsection*{\textsc{PatchNet} results: Fisher information}
\begin{figure}[htbp]
    \centering
    \includegraphics[width=0.5\linewidth]{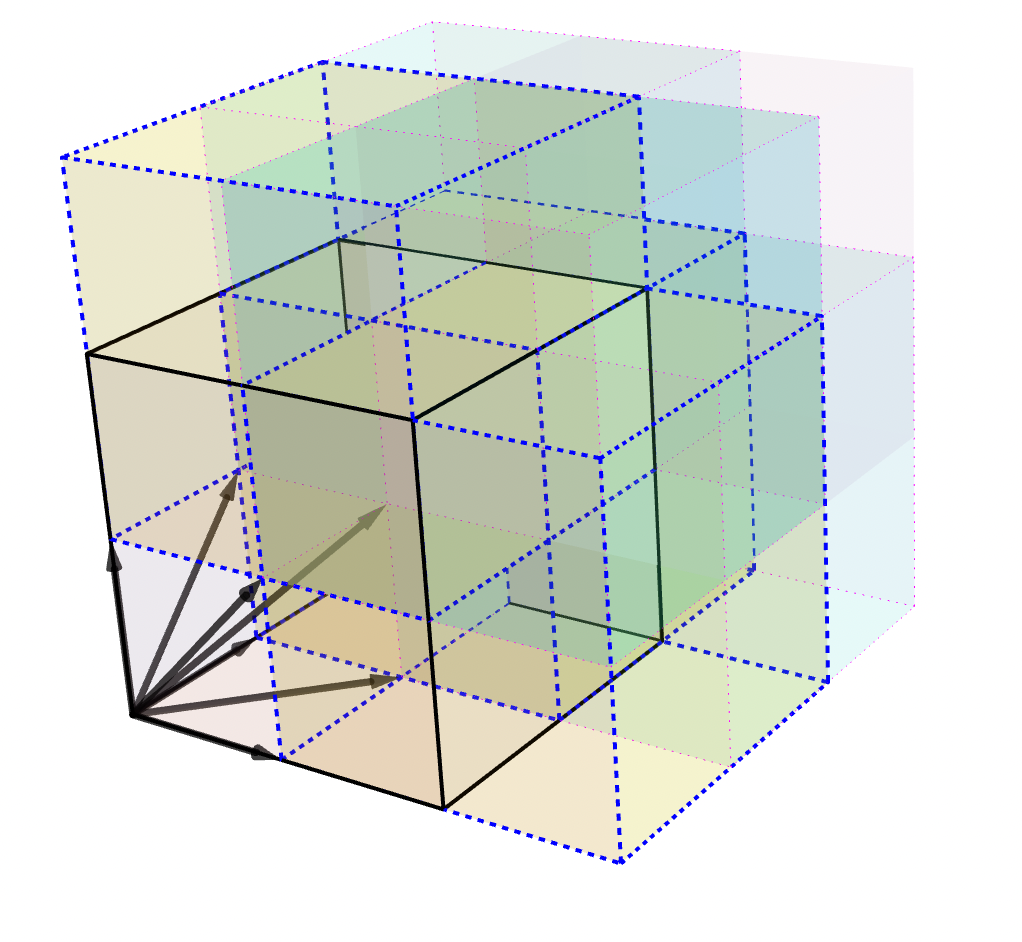}\includegraphics[width=0.5\linewidth]{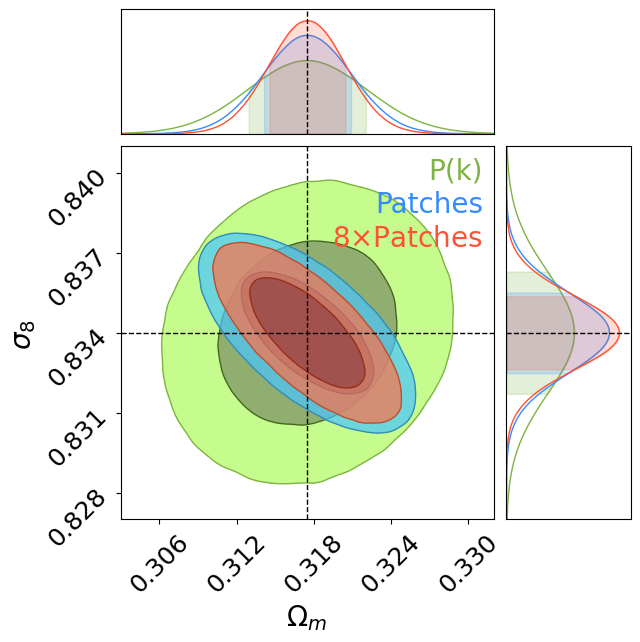} 
    \caption{Left: Overlapping patches to capture clustering signal straddling two adjacent patches. To include these information the patches are moved by half a patch width along 3 axes, 3 face diagonal and 1 body diagonal. We use the mean summary from these additional 7 patches, along with the original patch, during inference. Right: Information content in Patches. \textsc{PatchNet} trained on the $8^3$ discrete patches of $(125~\mathrm{Mpc}/h)^3$ volume extracts informative small-scale neural summary than the 2-point statistics. During inference, the use of overlapped patches ({8$\times$Patches}) further reduces the uncertainties of the parameters.}
    \label{fig:patchcorner}
\end{figure}

Figure \ref{fig:patchcorner} shows parameter uncertainties from the Fisher matrix of the mean estimate of these individual patch summaries ($\hat{\theta}:\{\Omega_m, \sigma_8\}$) which itself shows enhanced information content than the combined information from power spectra and bispectrum.\\
To avoid missing information in structures that cross the boundaries of patches we include an additional set of 7  patch sets each shifted by half a patch width (one for each of the $x$, $y$, and $z$ coordinates, one for each of the 3 face diagonals and one for the body diagonal)  for a total of 8. Since these patches provide a 8-fold cover of the simulation volume the information in these patches will be correlated. So as not to overcount information, this correlation is taken into account by  aggregating the parameter vectors from all patches before measuring the covariance matrix that goes into the Fisher information.
% \begin{figure}[htbp]
%     \centering
%     \includegraphics[width=4.35in]{fig/Pk+Bk+Patch.png}
%     \caption{Fisher from pk bk patch level summaries}
%     \label{fig:timeseries_para}
% \end{figure}
% \newpage
\section{Hierarchical scale-dependent  information content of dark matter density}
\label{hierarchicalinfo}
We evaluate the total Fisher information recoverable from combinations of global and local summaries of the matter density field. In particular, we analyze:
(i) the combination of the power spectrum $P(k)$ and patch-based neural summaries, and
(ii) an extended combination including the bispectrum $B(k)$, i.e., $P(k)+B(k)+\text{patches}$.

Our approach explicitly leverages the scale hierarchy in structure formation, combining interpretable 2-point and 3-point summaries with informative neural encodings of small-scale information in patches.   This hierarchical approach exploits scale complementarity: while $P(k)$ and $B(k)$ capture long-wavelength linear and quasi-linear modes, neural summaries of sub-volumes efficiently encode small-scale, nonlinear structures inaccessible to low-order correlation functions.

\begin{figure}[htbp]
    \centering
    \includegraphics[width=0.5\linewidth]{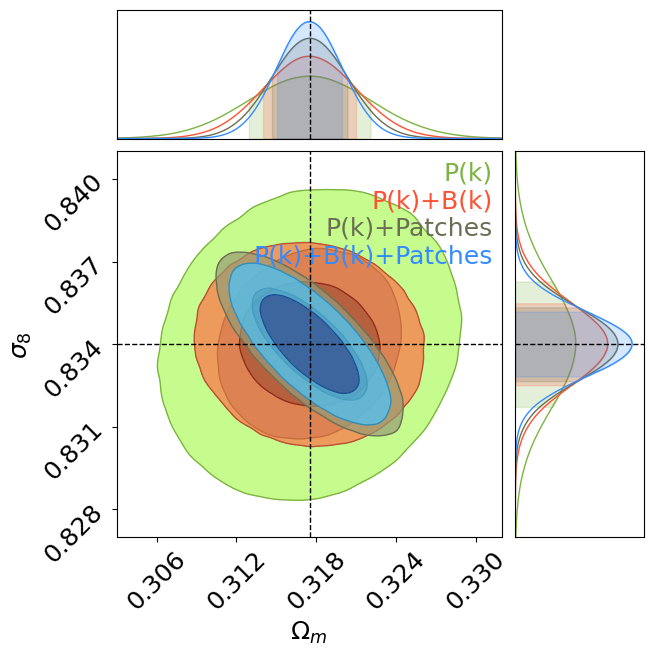}\includegraphics[width=0.5\linewidth]{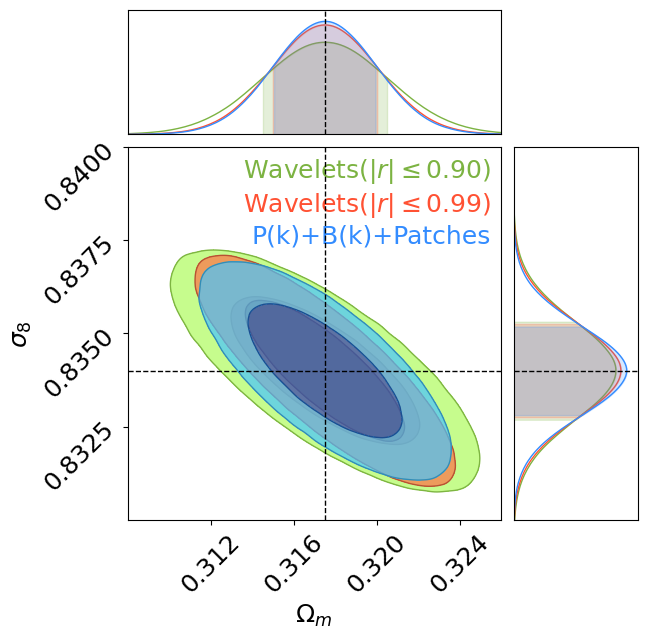} 
    \caption{Left: Cosmological constraints improve with scale hierarchy. $P(k)$, $B(k)$ contains information from linear and quasi-linear regimes, but fails to fully capture the non-gaussian structure. Thus, neural summary extracted from the Patches attempts to capture these non-linear information. The patch summary along with $P(k)$ and $B(k)$ give enhanced information. Right: Hierarchical information content vs Wavelets statistics. Wavelet information saturates at $|r|\leq0.99$, similar to information $P(k)+B(k)+$patches can extract. The convergence of information from two different methods supports the fact that hierarchical summaries can exploit the full information of \textit{Quijote} dark matter field.}
    \label{fig:patchvswst}
\end{figure}

The corner plot in Fig.~\ref{fig:patchvswst} shows the marginalized Fisher contours for each combination. We find that adding patch-based summaries to $P(k)$ significantly tightens parameter constraints, and including $B(k)$ further improves them. The full combination $P(k)+B(k)+\text{patches}$ recovers substantially more information than either component alone. Notably, it matches or outperforms wavelet scattering transforms (WST), a leading multi-scale summary method (\textit{cf.} Appendix \ref{Appendix:WST}).

While we cannot prove that we have reached the information limit, we arrive at very similar measures of information from two very different approaches, WST and  our hierarchical, hybrid strategy. We interpret this as evidence that we are approximating the full information content of our data set: a cubic $\ Gpc/h$ of dark matter represented on a grid of $128^3$ voxels.

\section{Discussion and conclusions}
\label{sec:conclusion}

We propose a field-based analysis as a more effective strategy for extracting cosmological information. Specifically, we introduce a hierarchical approach that integrates field-level information about small-scale structures from sub-volumes (patches) of the density field with the 2-point and 3-point correlations captured by the power spectrum and bispectrum. The power spectrum and bispectrum are known to be sufficient statistics in the perturbative, large-scale regime but fail to capture the full information content in the non-linear regime. By combining them with a field-based summary aggregated over patches---sensitive to small-scale, non-linear structures but large enough so the fundamental mode of each patch is in the linear regime---we have demonstrated that we can bridge this gap. 

Our hybrid approach has the potential to recover most, if not all, of the cosmological information by capturing both global and local features of the density field. As we demonstrate, this hybrid strategy significantly enhances the Fisher information content compared to using $P(k)$, $B(k)$ alone, and matches or exceeds the information in wavelet scattering transform coefficients, while remaining computationally tractable.

This work is conceptual in nature and focuses on the idealized case of dark matter (DM) fields, which are not directly observable. By working directly with DM, we avoid addressing the complex and unresolved problem of bias, \textit{i.e.,} the systematic differences between observable tracers (such as galaxies) and the underlying dark matter distribution, particularly in the non-linear regime \citep{Desjacques2018, Eisenstein_1998}. This does not mean that we set ourselves a trivial task since the dark matter field contains far more information than sparsely sampled halo or galaxy catalogs. 

Our primary aim is to find the highest possible lower bound on the cosmological information accessible within a $(1~\mathrm{Gpc}/h)^3$ volume of non-linearly evolved dark matter at the resolution sampled on $\sim$7.8 Mpc$/h$ voxels. This will be a lower bound since we cannot exclude that our inference approach is still somewhat suboptimal. However, the fact that we find nearly the same answer using an overcomplete set of wavelet scattering transfrom coefficients suggests that the bound at least no excessively loose. Studying the scaling of this information bound as a function of resolution and the extension of the \textsc{PatchNet} framework to more observationally relevant fields-such as the halo or galaxy distribution—is deferred to future work.

From a practical point of view, analyzing full volume, high-resolution, full-field data at the scale of current surveys monolithically on current GPUs is nearly infeasible due to memory and computational limitations, as well as the lack of scalable models capable of extracting parameters directly from raw fields. Moreover, generating simulations of such large volumes for training purposes is computationally expensive, further limiting the feasibility of full-field approaches. By contrast, our approach  enables sufficient training data to be extracted from a smaller number of simulations. 

When compared to  the wavelet scattering transform (WST), our patch-based strategy offers several additional potential practical advantages when applied to real world data: computing the WST on large volume, high resolution surveys with non-trivial geometries is computationally expensive and produces high-dimensional summaries that would require the additional step of finding lower dimensional embeddings. 
The \textsc{PatchNet} approach naturally extends to irregular survey geometries which can be treated by dividing them into patches and then aggregating. 

In addition, the WST is explicitly constructed to represent the spatial statistics of homogeneous data (whose statistics do not change as a function of position). Current and future surveys cover a significant redshift range and non-linear clustering is known to evolve as a function of redshift. \textsc{PatchNet} can be trained on patches in various redshift ranges, facilitating the application of neural field-based inference on light-cone data. 

Since \textsc{PatchNet} outputs a meaningful parameter vector for each density patch, a future version trained on galaxy data could be used to construct  three-dimensional parameter maps from galaxy surveys. This would be similar in spirit to the construction of parameter maps on the sphere  in \cite{2018JCAP...01..042M} which used fast, nearly optimal quadratic estimators to make such maps for nearly Gaussian fields on the two-dimensional sphere. Potential applications include exploring galaxy survey data for observational systematics or statistical anomalies.

We plan to explore these applications in future work.

\acknowledgments
AB acknowledges support from the Simons Foundation as part of the Simons Collaboration on Learning the Universe. The Flatiron Institute is supported by the Simons Foundation. All other post-analysis and neural training have been done on IAP’s Infinity cluster. BDW acknowledges support from the DIM ORIGINES 2023 INFINITY NEXT grant. OpenAI's language model -- ChatGPT was used for some language editing during the draft stage of the preparation of this manuscript.

\appendix
\section{Wavelet Scattering Transform }
 \label{Appendix:WST}
In this appendix we define the wavelet scattering transform (WST) and provide the details of our implementation, including the our postprocessing of the WST coefficients to avoid numerical instabilities in the computation of the WST Fisher information.

\subsubsection*{Solid Harmonic Wavelets}
The WST decomposes a given field using a sequence of wavelet convolutions, modulus operators, and local averaging, effectively hierarchically capturing multiscale structures and interactions. Wavelets are a family of localized oscillating waveforms dilated to different scales with zero mean. Here we use solid harmonic wavelets $\psi_l^m(\bold{r})$ which are specifically designed for analyzing three-dimensional data in a rotationally invariant manner 
\begin{equation}
    \label{mw}
    \psi_l^m(\bold{r})=\frac{1}{(2\pi)^{3/2}}\exp{-|\bold{r}|^2/2}|\bold{r}|^l\bold{Y}_l^m(\hat{\bold{r}}) .
\end{equation}
They are constructed using solid harmonics, which are solutions to Laplace’s equation in spherical coordinates and serve as natural basis functions for representing 3D structures. In order to capture information from different scales, the mother wavelet \ref{mw} is dilated at the scale $2^j$ 
\begin{equation}
    \psi_{j,l}^m(\bold{r})=2^{-3j}\psi_l^m(2^{-j}\bold{r}) .
\end{equation}
Solid harmonic wavelets decompose data in both radial and angular components, allowing for an efficient multiscale representation of features while preserving important symmetries. These wavelets are particularly useful in applications such as cosmology, fluid dynamics, and molecular imaging, where isotropy and multiscale structures are key characteristics of the underlying data. In the context of wavelet scattering transforms, solid harmonic wavelets enable hierarchical feature extraction while maintaining rotation equivariance and stability to deformations, making them valuable for cosmological parameter inference and large-scale structure analysis \citep{SimBIG:2023gke, DES:2023qwe}.

We will now give the definitions of the scattering coefficients.

\subsubsection*{Zero-order scattering coefficient}
It is defined as the integral of the input field $\rho(\bold{r})$ raised by some integral power $q$, effectively capturing the zeroth-order moment of the data distribution. Mathematically, it is expressed as 
\begin{equation}
    S_0[q]\rho(\bold{r})=\int \rho(\bold{r})^q d^3\bold{r} .
\end{equation}
As it doesn't contain any scale information, it is often interpreted as a measure of global statistical properties of the field.

\subsubsection*{First-order scattering coefficient}
The first-order wavelet scattering transform $(S_1)$ captures multiscale structural information by decomposing the input field using a family of wavelets indexed by scale $j$ and angular frequency $l$. The input field is convolved with the dilated wavelet $\psi_{j,l}^m(\bold{r})$ of given scale $j$ and angular frequency $l$ followed by taking Euclidean norm over index $m$ which ensures translational and rotational invariance of the coefficients respectively.
\begin{equation}
    U[j,l]\rho(\bold{r})=\left(\sum_{m=-l}^l|\rho(\bold{r})\star\psi_{j,l}^m(\bold{r})|^2\right)^{1/2} 
\end{equation}
The resulting transformed field $U[j,l]\rho(\bold{r})$ retains localized spatial structure while discarding phase information. The first-order scattering coefficient is then obtained by computing the zeroth-order moment of $U[j,l]\rho(\bold{r})$ raised to the integral power $q$ over the entire domain
\begin{equation}
    S_1[j,l,q]\rho(\bold{r})=\int |U[j,l]\rho(\bold{r})|^q d^3\bold{r} .
\end{equation}

\subsubsection*{Second-order scattering coefficient}
The second-order wavelet scattering coefficient($S_2$) extends the hierarchical feature extraction of the scattering transform by capturing interactions between structures at different scales and orientations. While $S_1$ describes localized amplitude patterns at individual scales, $S_2$ encodes correlations between first-order features, providing access to more complex, higher-order statistical information that is particularly useful in characterizing non-Gaussianity and filamentary structure in cosmological fields.

To compute $S_2$, the first-order modulus field $U[j,l]\rho(\bold{r})$ is further convolved with a second wavelet $\psi_{j^{\prime},l}^m(\bold{r})$, where $j<j^{\prime}$	ensures a hierarchical (coarse-to-fine) decomposition. Again, the Euclidean norm is taken over the azimuthal index $m$ to maintain rotational invariance
\begin{equation}
    U[j,j^{\prime},l]\rho(\bold{r})=\left(\sum_{m=-l}^l|U[j,l]\rho(\bold{r})\star\psi_{j^{\prime},l}^m(\bold{r})|^2\right)^{1/2}, \hspace{2cm} j<j^{\prime} .
\end{equation}
The second-order scattering coefficient is then defined as the spatial integral \citep{zhao2024simulationbasedinferencereionizationparameters} of the resulting field raised to an integer power $q$
\begin{equation}
    S_2[j,j^{\prime},l,q]\rho(\bold{r})=\int |U[j,j^{\prime},l]\rho(\bold{r})|^q d^3\bold{r} .
\end{equation}
\subsubsection*{Implementation}
We used \texttt{kymatio}~\citep{kymatio} to compute the wavelet scattering transform (WST) coefficients from the \textit{Quijote} dark matter overdensity field, rescaled as \((\delta+1)/2\). The decomposition was performed across scales \(j \in [0,6]\), corresponding to the maximum resolvable scale of the density field, with angular frequencies \(l \in [0,6]\) and modulus exponents \(q \in \{0.5, 1, 2, 3, 4\}\). For each combination of \(l\) and \(q\), the number of first-order coefficients is \(j_{\max}+1=7\), while the number of second-order coefficients is \(j_{\max}(j_{\max}+1)/2=21\). This configuration yields $5+(21+7)*7*5= 985$ coefficients in total for each realization.  To mitigate numerical instabilities caused by the large dynamic range of the coefficients, we normalize each WST coefficient by its standard deviation across the fiducial simulations.
\subsubsection*{Dimensionality reduction}
The wavelet scattering transform gives a high-dimensional vector consisting of highly correlated coefficients, with a great deal of redundancy. Combined with finite numerical precision, this causes the inverse of the covariance matrix to be numerically ill-defined, leading to failure to estimate an accurate FIM. To solve this issue, we regularize the covariance matrix by eliminating redundant scattering coefficients. We flag a pair of coefficients as redundant when it is highly correlated specifically when its  Pearson's $|r|$ exceeds $0.99$\citep{kendall1977advanced}. We only retain only one WST coefficient of each redundant pair in the WST feature vector. After this step 246 WST coefficients remain and the covariance matrix can be inverted. We show in Fig.~\ref{fig:patchvswst} the parameter constraints from the Fisher matrix, Eq.~\ref{infoineq}, calculated from this reduced set of wavelet summaries. To test the sensitivity of our FIM estimates to our choice of $r$ cutoff we  further remove coefficients relaxing our redundancy criterion to  $|r|>0.90$. While this  reduces the remaining number of WST coefficients to just 74, the FIM changes only slightly, as shown right hand panel of Figure \ref{fig:patchvswst}. This gives us confidence that our regularization of the WST Fisher matrix computation is robust. We use the WST coefficients with $r\leq0.99$  in what follows.

\section{Details on the neural estimator implementation}
\label{app:NN}
We explored around 20 different CNN architectures with different kernel sizes and hidden layers of various dimensions. Each of these models was then tested for both average pooling and max pooling separately after each convolution. We tried both using batch-normalization and without batch-normalization, and different activation functions, including sigmoid, tanh, softplus, ReLU, LeakyReLU as well as custom-made activation functions \cite{Bairagi:2025sux} for every model. We used popular models like ResNet18 and Inception net as well. These models were tested under various data-preprocessing conditions, \textit{i.e.}, rotation, flip, log transformation and true field and for different loss functions like mean squared error (MSE) and Log MSE loss. Several hundred models with different configurations were trained on the $1~\mathrm{Gpc}/h$ field before deciding on the full field CNN used in \ref{sec:fullfield}.

We tried approximately 10 different models with different hidden dimensions including Inception net for \textsc{PatchNet} architecture search. Similarly, the effect of different nonlinear activations, cost functions and batch-normalization  were tested for these models. We varied batchsize and number of patches to use during training to select the hyperparameters. We tested hundreds of combinations to determine the \textsc{PatchNet} architecture. 

As part of finding an aggregation method for information from different patches, we explored different weighting schemes instead of doing mean aggregation in the first place. We trained a FishNet \citep{fishnet} to get the covariance of each patch and  inverse-covariance-weight  the \textsc{PatchNet} output. Another effort of learning a weight for each patch was made to do exponential weighting on the patch level summary.
To aggregate information from the set of patch summaries we trained several MLPs on the combined summary of $P(k)$ and the aggregated patch summary from various models mentioned above. The patch summary was chosen as either final output or the output from the pre-final layer. In the end, these attempts did not extract more information than simply concatenating $P(k)$ and the mean-aggregated patch summary.

In total, we tested around 400-500 combinations of the above choices before finalizing our architecture and hyperparameter choice.

\bibliographystyle{JHEP}  % Numbered in order of citation  
\bibliography{final} 

\end{document}